\documentclass[twocolumn,twocolappendix]{aastex631}

\newcommand\blfootnote[1]{%
  \begingroup
  \renewcommand\thefootnote{}\footnote{#1}%
  \addtocounter{footnote}{-1}%
  \endgroup
}

\begin{document}

\title{Hunting Very High-Energy ($>$100 GeV) Emitting High-Synchrotron Peaked Blazars}

\author[0000-0001-5507-7660]{Sushmita Agarwal$^{*}$}
\affiliation{Department of Astronomy and Astrophysics, Tata Institute of Fundamental Research, Mumbai 400005, India}
\affiliation{Department of Astronomy, Astrophysics and Space Engineering, Indian Institute of Technology Indore, Khandwa Road, Simrol, Indore,
453552, India}

\author[0000-0001-7774-5308]{Vaidehi S. Paliya}
\affiliation{Inter-University Centre for Astronomy and Astrophysics (IUCAA), SPPU Campus, 411007, Pune, India}







\begin{abstract}

 Very-high energy (VHE; $>$100 GeV) $\gamma$-ray emission originates via some of the most extreme particle acceleration processes in the universe. Considering beamed active galactic nuclei, i.e., blazars, only a small fraction, mainly high synchrotron peak BL Lacs, have been detected in the VHE band with the ground-based Cherenkov telescopes. We utilized $\sim$16 years of Fermi-Large Area Telescope (LAT) observations in the 0.1$-$2 TeV energy range to systematically search for potential VHE emitters in a sample of high synchrotron peaked ($\nu^{\rm peak}_{\rm syn}>10^{15}$ Hz) BL Lac sources. We identified, for the first time, 92 VHE emitting blazars at $\geq 5\sigma$ confidence level. A significant VHE emission was also detected from 52 sources previously reported as VHE blazars. 
 Comparing with the general blazar population, these VHE emitting blazars are found to be located at low redshifts (mean $z=0.2 \pm 0.1$) and exhibit bright synchrotron emission ($\log F^{\rm peak}_{\rm syn}=-11.2 \pm 0.4$, in erg cm$^{-2}$ s$^{-1}$). We also investigated the coincidence of VHE photon arrivals with the source activity states and found that Fermi-LAT has detected VHE photons during both quiescent and elevated activity epochs. These VHE emitting blazars represent promising targets for current and next-generation ground-based Cherenkov telescopes, and provide powerful laboratories for probing particle acceleration in relativistic jets, testing multi-messenger connections, and constraining extragalactic background light models.

\end{abstract}



\section{Introduction} \label{sec:intro}
\blfootnote{$^{*}$Email: \href{sush.agarwal16@gmail.com}{sush.agarwal16@gmail.com}}
Blazars are a subclass of radio-loud active galactic nuclei (AGN) distinguished by relativistic jets oriented closely along the line of sight to the observer \citep{Urry1995PASP, Blandford2019ARA&A}. This alignment leads to strong Doppler boosting of the jet emission, resulting in brightness enhancement and extreme variability across the entire electromagnetic spectrum, with timescales ranging from years down to minutes. Their broadband spectral energy distributions (SEDs) in $\nu\,-\nu\,f_{\nu}$ representation exhibit a characteristic double-humped structure, indicative of two distinct non-thermal emission processes. The low-energy hump, extending from the far-infrared to the soft X-ray regime, is attributed to synchrotron radiation from relativistic electrons in the jet. The high-energy component, extending from hard X-rays to $\gamma$-rays and occasionally into the TeV range, is generally ascribed to inverse Compton scattering of relativistic electrons by photons internal or external to the jet or to hadronic emission processes \citep{Abdo2010b, Maraschi1992ApJ, Dermer_2009, Sikora1994ApJ, Mannheim1993A&A, Rachen_1999,Mucke2001APh}.

The synchrotron peak in the SED varies across blazars and serves as a key indicator of the jet’s particle acceleration efficiency \citep{Giommi2012A&A}. Based on the synchrotron peak frequency ($\nu^{\rm peak}_{\rm syn}$), blazars are typically categorized into three subclasses: low-synchrotron-peaked (LSP; $\nu^{\rm peak}_{\rm syn} < 10^{14}$ Hz), intermediate-synchrotron-peaked (ISP; $10^{14} < \nu^{\rm peak}_{\rm syn} < 10^{15}$ Hz), and high-synchrotron-peaked (HSP; $\nu^{\rm peak}_{\rm syn} > 10^{15}$ Hz) blazars \citep[][]{Abdo2010b}. A subset of HSP sources visibly showcase synchrotron peaks extended into the keV regime -- commonly referred to as extreme blazars (EHSP) \citep[$\nu^{\rm peak}_{\rm syn} > 10^{17}$ Hz; see, e.g.,][]{Ghisellini1999APh,2001A&A...371..512C,2019MNRAS.486.1741F}. HSP and EHSP BL Lac objects stand out for their characteristically hard $\gamma$-ray spectral profiles, effective particle acceleration mechanisms, and frequent detection in the very high-energy (VHE; E $\geq$100 GeV) $\gamma$-ray band \citep[cf.][]{2019ApJ...882L...3P,Biteau2020NatAs}. 

Observationally, HSPs and EHSPs are prime candidates for VHE observations. This arises because relativistic electrons in HSPs are accelerated to higher energies than in LSPs, enhancing the likelihood of inverse Compton upscattering to produce VHE $\gamma$-rays \citep{Stecker1996ApJ, Sambruna1996ApJ, Costamante2002A&A, Sergio_Colafrancesco_2006, Weekes2008AIPC,Arsioli2017A&A}. An updated list of known VHE detected sources, many of which are HSPs, is maintained in the web-based TeVCat\footnote{\url{https://www.tevcat.org/}} catalog \citep{Wakely2008ICR}. BL Lac objects with radiative output upto TeV energies serve as powerful probes of extreme astrophysical environments, offering insights into particle acceleration mechanisms \citep{Boettcher2007, Saugé_2004}, potential associations with high-energy neutrinos and cosmic rays \citep{Murase2014PhRvD,SurayMNRAS2023,Chang2022ApJ,IceCube2018Sci,Das2020ApJ}, and enabling measurements of the extragalactic background light (EBL) and star formation rate \citep{Biteau_2015, GiommiMNRAS2015,Aharonian2006Natur,Finke2022ApJ} and constraints on key aspects of $\gamma$-ray cosmology, such as the intergalactic magnetic field \citep{Neronov2010Science,Finke_2015,Aharonian_2023}. 

Identifying potential VHE $\gamma$-ray candidates has become increasingly important with the rise of multimessenger astrophysics and the improved sensitivity expected from the next generation of VHE observatories for a more targeted survey of sky. However, current VHE astronomy is constrained by a limited number of known TeV-emitting sources -- as reflected in the relatively sparse entries of the TeVCat catalog and by observational challenges inherent to imaging atmospheric Cherenkov telescopes (IACTs) such as Major Atmospheric Gamma Imaging Cherenkov (MAGIC) telescope, Very Energetic Radiation Imaging Telescope Array System (VERITAS), High Energy Stereoscopic System (H.E.S.S.), and Whipple observatory. These instruments usually have low-duty cycle, as observations can only be conducted during clear, moonless nights, limiting data collection to roughly 10–15\% of the year. Additionally, weather conditions like humidity and atmospheric instability can degrade data quality. IACTs also have a narrow field of view, restricting their ability to conduct wide-area surveys and necessitating targeted monitoring strategies \citep{HESS2025A&A}. 

In contrast, the Fermi Large Area Telescope (Fermi-LAT) has been performing continuous all-sky monitoring since 2008 in the energy range of 100 MeV to 2 TeV. Although its sensitivity decreases at higher energies, it benefits from a large field of view and continuous exposure, which helps mitigate background contamination and makes it well-suited for identifying promising candidates in the VHE domain ($E>100\,\rm{GeV}$). The overlap between Fermi-LAT's upper energy range and the VHE regime offers a complementary approach for uncovering new VHE-emitting blazars, thus providing potential targets for follow-up with current and future ground-based VHE instruments \citep[see, e.g.,][]{2026JHEAp..4900454T,2025ApJ...991L...8P}.

In this work, we take advantage of Fermi-LAT's high-energy sensitivity and its extensive all-sky monitoring since 2008 to search for potential VHE-emitting blazars. Given that HSP blazars are the best candidates for VHE detection, we use the population of bonafide HSPs as a seed sample for further investigation. The strategy for candidate identification and sample selection is outlined in Section~\ref{sec:sample_selection}. Details of the data reduction and analysis procedures are provided in Section~\ref{sec:fermi_analysis}. The list of newly identified candidates is presented in Section~\ref{sec:new_candidates}, and the results of the population-level study are discussed in Section~\ref{sec:discussion}.

\section{Sample selection}  \label{sec:sample_selection}
The fourth data release of the Fermi-LAT $\gamma$-ray source catalog includes 7194 objects detected over 14 years of observations in the 50 MeV to 1 TeV energy range \citep[4FGL-DR4;][]{2022ApJS..260...53A,2023arXiv4FGL_DR4}. 
A subset of these sources are recognized as VHE emitters, with counterparts listed in the TeVCat online catalog \citep{Wakely2008ICR}. Currently, the identification of VHE sources largely depends on follow-up observations triggered by detections from ground-based and space-based multi-wavelength facilities. Given the continuously growing Fermi-LAT dataset and its all-sky monitoring capabilities, it presents a valuable instrument to identify new VHE candidates and support future efforts in VHE $\gamma$-ray astronomy. 

Among the most promising VHE source classes are HSP blazars \citep{Arsioli2015A&A}, which are considered the best targets for follow-up VHE observations \citep{Chang2017A&A, Arsioli2017A&A}. The most comprehensive and up-to-date sample of these sources is provided by the 3HSP catalog, which includes 2013 HSP blazars \citep{Chang2019A&A}. 

The primary objective of this work is to identify potential VHE-emitting sources among $\gamma$-ray detected HSP sources, using the 4FGL-DR4 catalog. To achieve this, we cross-matched the 4FGL-DR4 and 3HSP catalogs using a search radius of 5 arcseconds, leading to the selection of 1004 HSP sources. Additionally, recognizing that about 30\% 4FGL-DR4 sources lack multiwavelength counterparts, we also considered 41 3HSP blazars that lie within the 95\% positional uncertainty regions of unassociated Fermi sources, and hence likely to be the low-frequency counterpart of the unassociated $\gamma$-ray sources. This results in a parent sample of 1045 $\gamma$-ray emitting 3HSP sources, which we further studied for signatures of VHE emission.

\section{Data Reduction} \label{sec:fermi_analysis}
We utilized $\sim$16 years of Fermi-LAT observations conducted from MJD 54683 to 60587 (2008 August 5 to 2024 October 4). We performed the standard binned likelihood method using fermiPy \citep{Wood:2017TJ} with fermitools (v 2.2.0). The data was analyzed using the latest instrument response function {\tt\string P8R3\_SOURCE\_V3} in the energy range 100 GeV$-$2 TeV and utilized photons within $2^{\circ}$ around the source location, considering the PSF of the telescope above 100 GeV \citep[$\sim$0.1$^{\circ}$;][]{2021ApJS..256...12A}. The {\tt \string SOURCE} class events (evclass$=$128) were spatially binned with $0.05^{\circ}$ per pixel and divided into ten logarithmically spaced bins per energy decade. To access the robustness of detections against potential cosmic-ray contamination, we also repeated the analysis with progressively stricter LAT event selections: CLEAN (evclass$=$256) and ULTRACLEAN (evclass$=$512) classes. Further, to ensure high-quality data corresponding to the good time intervals, we employed the filter expression recommended by the LAT team {\tt\string DATA\_QUAL} $>$ 0 \&\& {\tt\string LAT\_CONFIG==1}. Solar disk is expected to be a significant contributor for the VHE emission \citep{Albert2023PhRv}. To correct for potential contamination from solar-disk $\gamma$-ray emission \citep[e.g.,][]{Arsioli2024ApJ,Linden2020arXiv}, we applied an additional data-cleaning step by excluding time intervals during which the source position was within $2^\circ$ of the Sun, using the filter
{\tt\string angsep($\rm{\tt \string RA_{ SOURCE}}$, $\rm{\tt \string DEC_{SOURCE}}$, $\rm{\tt \string RA_{SUN}}$, $\rm{\tt \string DEC_{SUN}}$) $>$ 2}.
This ensures that transient solar activity does not influence the VHE signatures studied in this work. We also adopted a zenith angle cut of $105^{\circ}$ to eliminate contamination from secondary $\gamma$-rays from Earth's albedo. An initial optimization is performed at this stage to create an initial estimate for the model for the region.

The cleaned events data was then modeled by considering all 4FGL-DR4 sources lying within $2^{\circ}$ of the source location. Considering the low photon statistics, the $\gamma$-ray emission from the source of interest was modeled with Power-Law 2 model\footnote{\url{https://fermi.gsfc.nasa.gov/ssc/data/analysis/scitools/source_models.html}} parametrized as :
\begin{equation}
\frac{dN}{dE}=\frac{N\,(\Gamma + 1)\,E^{\Gamma}}{E^{\Gamma+1}_{\rm max} - E^{\Gamma+1}_{\rm min}}    
\end{equation}
where N is integral flux, over the range $E_{\rm min}=100\, \rm{GeV}$ to $E_{\rm max}=2\,\rm{TeV}$ and allowed to vary between initial optimized flux values factored by 1000. The spectral index parameter $\Gamma$ was set free and allowed to vary between 0 to 10. Bright sources with Test statistics (TS) $>25$ were allowed to vary, whereas sources with TS$<1$ were removed from the model. We additionally accounted for contributions from galactic and isotropic background by using {\tt\string gll\_iem\_v07} and {\tt\string iso\_P8R3\_SOURCE\_V3\_v1} template \citep[][]{2016ApJS..223...26A}. We also computed the time of arrival of photons having more than $95\%$ probability of association with the source using the tool {\tt \string gtsrcprob}. For sources with TS$>25$, we repeated the likelihood fitting using the EBL model provided by \citet{2011Dominguez} to estimate the intrinsic, i.e., EBL-attenuation corrected, spectral parameters.

\section{Results}\label{sec:new_candidates}
The Fermi-LAT data analysis of 1045 sources led to the identification of 144 $\gamma$-ray sources detected at $\gtrsim 5\sigma$ confidence level, i.e., TS$\,\geq\,$25. Additionally, 116 sources were detected with $12\leq{\rm TS}<25$. Cross-referencing the parent sample of 1045 objects with the TeVCat, we found that 60 of them have already been reported as VHE emitters in previous works. Of these, 52 are significantly detected in our analysis with TS $\geq $25, while 7 sources have $12<{\rm TS}<25$ (4FGL J1103.6$-$2329, 4FGL J1518.0$-$2731, 4FGL J1230.2+2517, 4FGL J1442.7$+$1200, 4FGL J1958.3$-$3010, 4FGL J0847.2$+$1134 , and 4FGL J0013.9$-$1854). The only known VHE emitting blazar missing in our list is the HSP source 4FGL J0152.6+0147 (associated with RGB J0152+017, $z=0.08$) with a TS of 2.5. The source was detected in VHE energy range only in 2007 October and November by H.E.S.S., i.e., before the launch of the Fermi satellite \citep{Aharonian2008A&A}. Therefore, it is likely that during the period of the Fermi-LAT operation, the source remained in quiescence and hence was not detected with the satellite. Therefore, we have identified 92 new VHE emitting HSP blazars using the $Fermi$-LAT data. Their spectral parameters estimated from the Fermi-LAT data analysis are provided in Table~\ref{tab:new_candidates_1}, whereas those for already known VHE emitters are listed in Table~\ref{tab:known_VHE_candidates}. Sources detected at low significance ($12\leq{\rm TS}<25$) are tabulated in the appendix (Table \ref{tab:ts_betwen_12_and_25}). 

To assess the impact of cosmic-ray–induced background on our sample, we re-analyzed all significantly detected sources (TS$>$25) using CLEAN and ULTRACLEAN event classes. Eleven objects fell below the detection threshold (TS$=$25) when using CLEAN events, while 19 with ULTRACLEAN events. There are seven blazars common in both analyses, giving a total of 23 sources ($\sim$16\%) that did not pass at least one of the stricter event-class cuts. A comparison of the number of associated photons and TS values for these 23 sources, relative to the SOURCE class, is provided in Table~\ref{tab:ultraClean_class_ts_lessthan25}. This modest reduction confirms that most detections are stable against changes in event-class selection, implying minimal contamination from residual cosmic-ray events.

\begin{deluxetable*}{lllllllllll}
\tablewidth{0pt}
\tablecaption{The spectral parameters of the newly identified VHE emitting blazars. \label{tab:new_candidates_1}}
\tablehead{
\colhead{Source Name} & 
\colhead{z} & 
\colhead{TS$_{\rm obs}$} & 
\colhead{F$_{\rm obs}$} & 
\colhead{$\Gamma_{\rm obs}$} & 
\colhead{TS$_{\rm EBL}$} & 
\colhead{F$_{\rm EBL}$} & 
\colhead{$\Gamma_{\rm EBL}$} & 
\colhead{No.} & 
\colhead{E$_{\rm max}$} & 
\colhead{$\rm Time_{\rm arr}$} \\
\colhead{[1]} & 
\colhead{[2]} & 
\colhead{[3]} & 
\colhead{[4]} & 
\colhead{[5]} & 
\colhead{[6]} & 
\colhead{[7]} & 
\colhead{[8]} & 
\colhead{[9]} & 
\colhead{[10]} &
\colhead{[11]}  
}
\startdata
J0014.7$+$5801 & $-$ & 27 & $0.77 \pm 0.50$ & $3.33 \pm 1.27$   & $-$ & $-$ & $-$ & 3 & 160.8 & 54957.0 \\
J0015.6$+$5551 & 0.217 & 39 & $1.57 \pm 0.83$ & $2.80 \pm 0.77$   & 40 & $1.53 \pm 0.74$ & $2.01 \pm 0.88$ & 4 & 230.0 & 58384.7 \\
J0022.0$+$0006 & 0.306 & 41 & $1.61 \pm 0.86$ & $3.34 \pm 1.06$   & 41 & $1.56 \pm 0.77$ & $2.36 \pm 1.24$ & 3 & 353.8 & 57494.7 \\
J0047.9$+$5448 & $-$ & 26 & $1.28 \pm 0.91$ & $2.16 \pm 0.79$   & $-$ & $-$ & $-$ & 4 & 605.9 & 56692.3 \\
J0051.2$-$6242 & 0.156 & 30 & $0.45 \pm 0.33$ & $8.35 \pm 4.07$   & 30 & $0.46 \pm 0.34$ & $8.08 \pm 4.28$ & 3 & 151.8 & 58968.8 \\
\enddata
\tablecomments{The column information are as follows: [1]: 4FGL source name; [2]: spectroscopic redshift of the source from \citet{Arsioli2025MNRAS} and \citet{2021ApJS..253...46P}. For sources with no robust spectroscopic redshift measurement a `--' entry has been made, [3], [4] and [5]: observed TS, energy flux (in 10$^{-6}$ MeV cm$^{-2}$ s$^{-1}$), and photon index, in the energy range of 0.1$-$2 TeV; [6], [7], and [8]: EBL attenuation corrected TS, energy flux (in 10$^{-6}$ MeV cm$^{-2}$ s$^{-1}$), and photon index, in the energy range of 0.1$-$2 TeV (only listed for sources with available spectroscopic redshift measurement); [9]: number of $>$100 GeV photons detected from the source having $\geq$95\% association probability;  [10]: energy of the maximum energy photon (in GeV), and [11]: time of arrival of the maximum energy photon (in MJD). The complete version of this table is available on ZENODO (doi: \href{https://doi.org/10.5281/zenodo.17593405}{10.5281/zenodo.17593405}). In the associated catalog file, entries marked `--' in the manuscript are replaced with `–999'.}
\end{deluxetable*}

\begin{deluxetable*}{lllllllllll}
\tablewidth{0pt}
\tablecaption{The spectral parameters of known VHE emitting blazars in TeVCAT. \label{tab:known_VHE_candidates}}
\tablehead{
\colhead{Source Name} & 
\colhead{z} & 
\colhead{TS$_{\rm obs}$} & 
\colhead{F$_{\rm obs}$} & 
\colhead{$\Gamma_{\rm obs}$} & 
\colhead{TS$_{\rm EBL}$} & 
\colhead{F$_{\rm EBL}$} & 
\colhead{$\Gamma_{\rm EBL}$} & 
\colhead{No.} & 
\colhead{E$_{\rm max}$} & 
\colhead{$\rm Time_{\rm arr}$} \\
\colhead{[1]} & 
\colhead{[2]} & 
\colhead{[3]} & 
\colhead{[4]} & 
\colhead{[5]} & 
\colhead{[6]} & 
\colhead{[7]} & 
\colhead{[8]} & 
\colhead{[9]} & 
\colhead{[10]} &
\colhead{[11]} 
}
\startdata
J0033.5$-$1921 & $-$ & 36 & $1.49 \pm 0.93$ & $3.19 \pm 1.14$  & $-$ &  $-$ &  $-$ & 3 & 170.0 & 54956.5 \\
J0035.9$+$5950 & 0.467 & 396 & $8.54 \pm 1.75$ & $2.97 \pm 0.34$   & 398 & $8.04 \pm 1.43$ & $1.11 \pm 0.44$ & 29 & 655.3 & 57381.2 \\
J0136.5$+$3906 & $-$ & 82 & $1.50 \pm 0.62$ & $4.29 \pm 1.22$   &  $-$ &  $-$ &  $-$ & 7 & 250.6 & 56683.5 \\
J0214.3$+$5145 & 0.048 & 61 & $2.62 \pm 1.08$ & $3.01 \pm 0.67$   & 61 & $2.61 \pm 1.05$ & $2.86 \pm 0.68$ & 5 & 179.2 & 58240.6 \\
J0232.8$+$2018 & 0.139 & 30 & $1.81 \pm 1.14$ & $2.53 \pm 0.82$   & 31 & $1.75 \pm 1.03$ & $2.00 \pm 0.89$ & 3 & 308.1 & 59508.4 \\
\enddata
\tablecomments{The column descriptions are identical to those in Table~\ref{tab:new_candidates_1}. The complete version of this table is available on ZENODO (doi: \href{https://doi.org/10.5281/zenodo.17593405}{10.5281/zenodo.17593405}). In the associated catalog file, entries marked `--' in the manuscript are replaced with `–999'.}
\end{deluxetable*}

\vspace{-1.6cm}
There are 456 HSP sources from which at least one VHE photon was detected by the Fermi-LAT. All 144 significantly detected objects (with TS $\geq$25) have been found to emit two or more VHE photons with $\geq 95\%$ association probability, thus robustly confirming them to be VHE emitting blazars \citep[Table~\ref{tab:new_candidates_1} and \ref{tab:known_VHE_candidates}, see also,][]{2013ApJ...777L..18T}. We also found 312 sources that emitted at least one VHE photon; however, their detection significance is low (TS $<$ 25). This includes 116 sources with TS values between 12 and 25, and 196 sources with TS $<$12. Though their VHE detection cannot be claimed at high significance, these objects can be considered as candidate VHE-emitting blazars. In the appendix, we provide the relevant VHE photon information (Table \ref{tab:ts_betwen_12_and_25} and \ref{tab:ts_less_than_9}).

Out of 41 unassociated $\gamma$-ray sources, four were detected with TS$\geq$9, and one VHE photon with $\geq$95\% association probability was detected from ten sources (see Table \ref{tab:unassociated_sources}). This finding suggests these HSP blazars to be the likely counterpart of the unassociated VHE emitting $\gamma$-ray sources.

VHE $\gamma$-rays traveling cosmological distances undergo pair production when interacting with EBL photons, leading to energy-dependent flux suppression that becomes increasingly significant at redshifts $z \gtrsim 0.1$, especially for $\gamma$-rays with energies above 100 GeV \citep[e.g.,][]{Hauser2001ARA&A}. We applied EBL absorption correction to recover the intrinsic spectra utilizing the spectroscopic redshift for the identified candidates \citep[e.g.,][]{Arsioli2025MNRAS,2021ApJS..253...46P}. EBL-corrected spectral parameters are provided in Tables~\ref{tab:new_candidates_1} and ~\ref{tab:known_VHE_candidates}. A significant fraction of these sources exhibit EBL-corrected photon indices below 2.0, indicating them to have intrinsically hard $\gamma$-ray spectrum \citep[see also,][]{2019ApJ...882L...3P}. 

Some sources exhibit extremely soft spectra ($\Gamma_{\rm obs}=10$), which likely result from the spectral fit reaching its upper boundary due to the spectrum being too steep or poorly constrained, typically because of low photon statistics. Two sources - J1243.2+3627 and J1340.5+4409 have TS$\geq$25, yet their best-fit photon indices reached the boundary value of $\Gamma=10$. Additionally, a few sources with TS values between 12 and 25 also reached the upper limit of the photon index. These cases are flagged in the \texttt{index\_Flag} column in the catalog for Table \ref{tab:ts_betwen_12_and_25}.

\section{Discussion} \label{sec:discussion}
The origin of VHE emission in blazars remains a relatively unexplored problem in high-energy astrophysics. Although the number of confirmed VHE blazars is limited, they offer valuable insight into extreme particle acceleration and the physical conditions that enable such high-energy processes. Expanding the sample of VHE candidates is particularly important in view of the upcoming Cherenkov Telescope Array Observatory (CTAO), which will greatly enhance detection sensitivity in this regime. A well-defined candidate list is therefore essential for guiding targeted observations and advancing our understanding of extreme particle accelerators. In this work, we report, for the first time, the identification of 92 VHE $\gamma$-ray emitting blazars and confirm VHE detections of 52 sources previously identified in this energy range. 

\begin{figure*}
    \centering
    \includegraphics[scale=0.3]{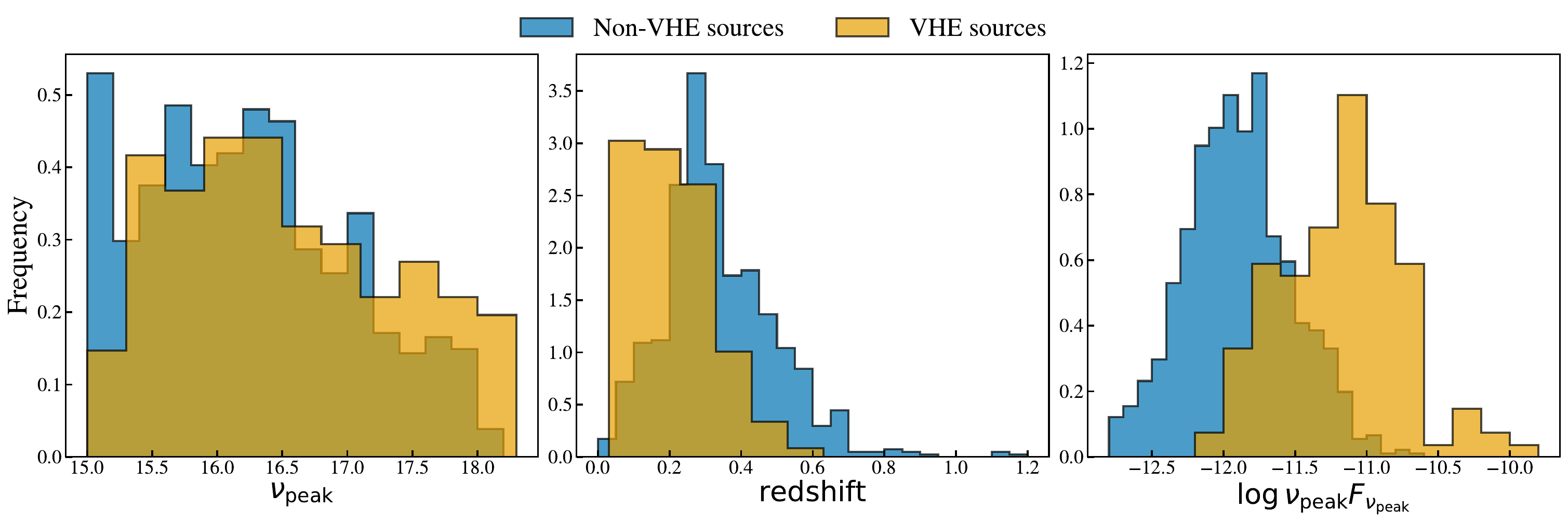}
    \caption{Left: Distribution of synchrotron peak frequency for the 4HSP-Fermi non-VHE population (blue) and VHE sources identified in this study (yellow). Middle: Redshift distribution for the same populations. Right: Distribution of synchrotron peak flux for the same populations.}
    \label{fig:Distribution_vpeak_z_nuFnu}
\end{figure*}

\subsection{Comparison with non-VHE Emitting HSP Blazars}
To understand the origin of the VHE emission, we compared the brightness, redshift, and synchrotron peak properties of the identified VHE sources and non-VHE-detected HSP sources using data from the 4FGL-DR4 catalog (see Figure~\ref{fig:Distribution_vpeak_z_nuFnu}). The synchrotron peak frequency distribution of the VHE emitters ($\log \nu^{\rm peak}_{\text{syn}} = 16.5 \pm 0.9$, in Hz) peaks at slightly higher frequency compared to the non-VHE 3HSP–Fermi sources ($\log \nu^{\rm peak}_{\text{syn}} = 16.1 \pm 0.8$, in Hz) though the dispersion is large. Furthermore, most of the VHE blazars are located in the nearby universe, with a mean redshift of $0.2 \pm 0.1$, which is lower than the non-VHE population ($0.3 \pm 0.2$), though with a large spread. Additionally, the synchrotron peak flux ($\log \nu F_{\nu,\text{syn}}$) for the VHE emitters is systematically higher, peaking at $-11.2 \pm 0.5$ (in erg cm$^{-2}$ s$^{-1}$), compared to $-11.9 \pm 0.4$ (in erg cm$^{-2}$ s$^{-1}$) for non-VHE sources, hinting at a potential brightness-driven bias in the VHE detection. This is because, considering one-zone leptonic models, a brighter synchrotron emission peaking at X-rays corresponds to the inverse Compton spectrum peaking at hundreds of GeV energies, making the source brighter at VHE energies. Indeed, the detection of hard $\gamma$-ray spectra at MeV-GeV energies from the HSP blazars supports this hypothesis \citep[e.g.,][]{2022ApJS..263...24A}.


\begin{figure}
\centering
\includegraphics[scale=0.35]{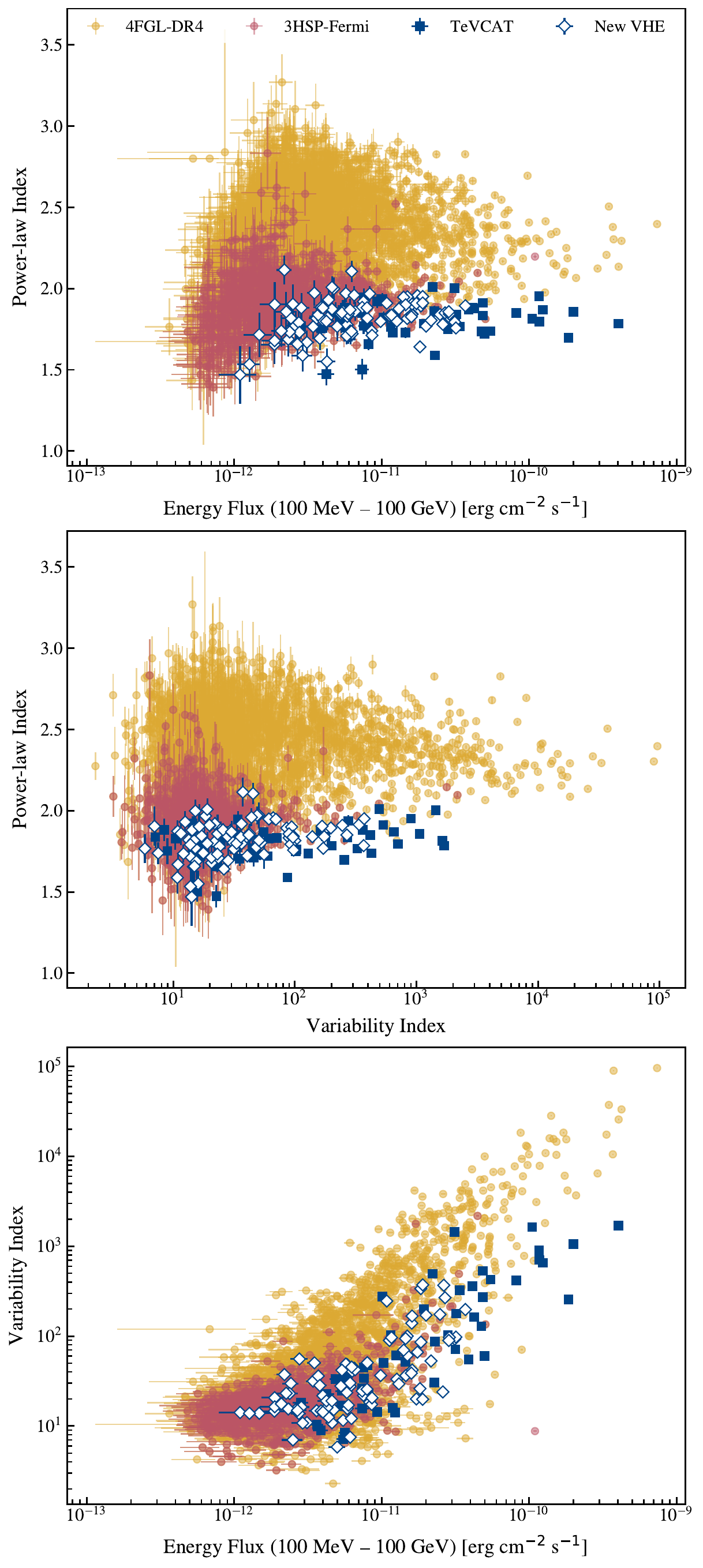}
\caption{Comparison of spectral and variability properties among different $\gamma$-ray blazar samples. Top: Power-law photon index versus energy flux (100 MeV–100 GeV) for 4FGL-DR4 blazars (orange circles), 3HSP–Fermi sources (red circles), previously known VHE emitters from TeVCat (filled blue squares), and newly proposed VHE candidates identified in this work (open blue diamonds; listed in Table~\ref{tab:new_candidates_1}). Middle: Photon index versus variability index for the same source populations. Bottom: Variability index as a function of energy flux for the same samples as in the upper panels.
}\label{fig:PLindex_energyflux_variability_index}
\end{figure}

\subsection{Comparison with General Gamma-ray Emitting Blazar Population}
In Figure~\ref{fig:PLindex_energyflux_variability_index}, we show the variations of the power law photon index, variability index, and energy flux with respect to each other, by adopting these values from the 4FGL-DR4 catalog. Such comparisons enable us to explore the distribution of VHE-emitting blazars within the spectral and variability parameter space and reveal clear distinctions from the broader $Fermi$-LAT detected blazar population. As illustrated in Figure \ref{fig:PLindex_energyflux_variability_index}, newly identified VHE blazars preferentially exhibit hard photon spectra (median photon index $1.8 \pm 0.1$) and moderate-to-high energy fluxes in the 0.1$–$100 GeV range ($\gtrsim 1.1 \times 10^{-12}\,\rm erg\,cm^{-2}\,s^{-1}$), whereas the general 4FGL-DR4 sample includes a substantial fraction of softer, fainter sources. Both newly identified VHE sources and already known VHE emitters share a similar parameter space in all the plots.

Interestingly, a majority of the VHE emitting sources tend to display relatively low-to-moderate variability indices (median value of $35$) for sources of comparable flux. This observation can be explained as follows: the detection of a hard $\gamma$-ray spectrum (0.1$-$100 GeV photon index $<$2) implies that the inverse Compton peak lies at energies $>$100 GeV. Therefore, the 0.1$-$100 GeV emission is mainly produced by the low-energy electron population whose cooling time is longer, hence lesser variability, compared to high-energy electrons producing $>$100 GeV to TeV radiation \citep[e.g,][]{1999MNRAS.306..551C,2020ApJS..248...29A}.

\begin{figure}
\centering
\includegraphics[scale=0.35]{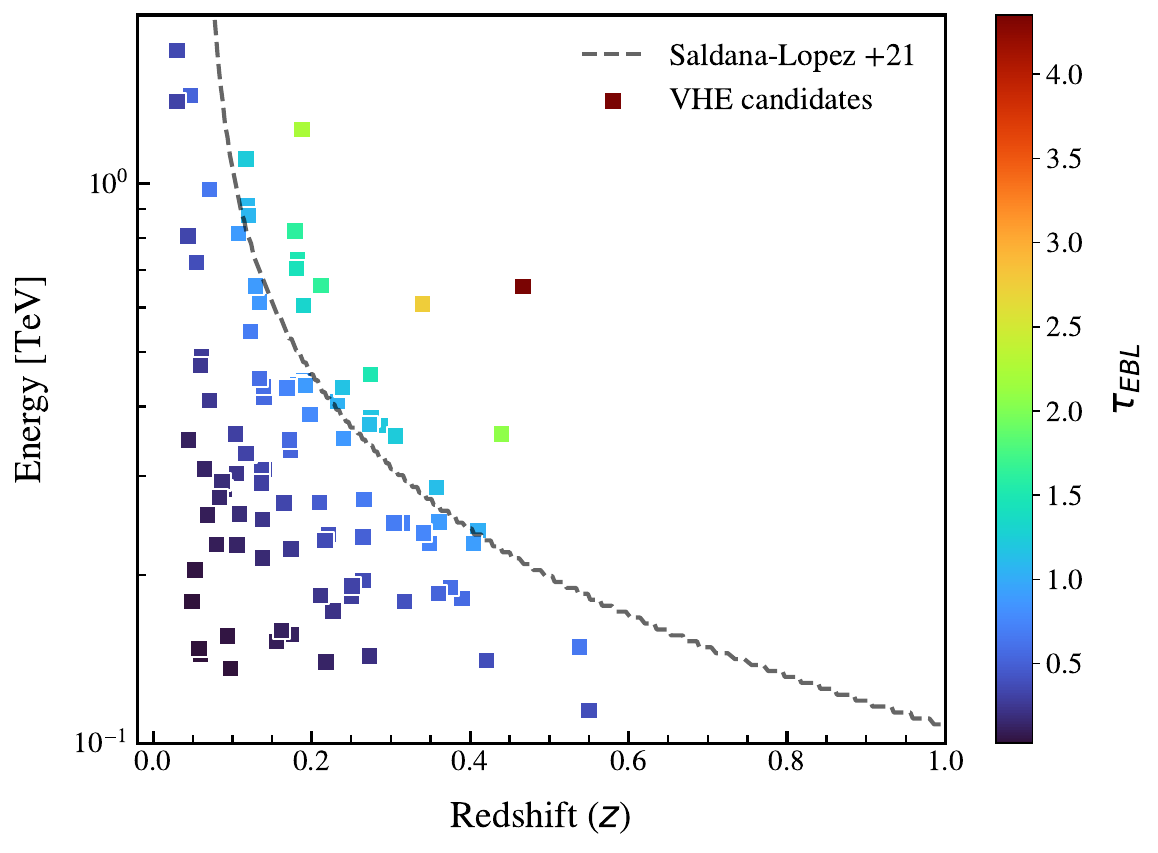}
\caption{Cosmic $\gamma$-ray horizon. Highest-energy photons detected from identified VHE-emitting HSP blazars, plotted as a function of their spectroscopic redshifts, based on the Fermi-LAT analysis in the 0.1–2 TeV energy range. The gray dashed line indicates the opacity regime corresponding to an EBL optical depth of $\tau_{EBL}=1$ \citep{Saldana-Lopez:2020qzx}. Source markers are color-coded by their respective EBL optical depth values derived from the same model.}
\label{fig:energy_redhsift_tauEBL}
\end{figure}

\vspace{2mm}
\subsection{Cosmic Gamma-ray horizon}
In addition to the jet energetics and spectral properties of the source, the detectability of VHE $\gamma$-ray sources is significantly influenced by attenuation due to the EBL, which limits $\gamma$-ray propagation over cosmological distances. To evaluate this effect for the newly identified VHE emitting HSP blazars, we plotted the energy of the maximum energy photon as a function of the blazar redshift in Figure~\ref{fig:energy_redhsift_tauEBL}. We also show the cosmic $\gamma$-ray horizon, i.e., EBL optical depth ($\tau_{\rm EBL}=1$) with a dashed line following the EBL model proposed by \citet{Saldana-Lopez:2020qzx}.

The majority of VHE emitting blazars lie below or near this horizon, indicating that they are situated in regions where the Universe remains transparent to VHE $\gamma$-rays. This reinforces their viability as detectable targets for current and future ground-based $\gamma$-ray observatories. The highest redshift source in our analysis -  4FGL J0506.0$+$6113 \citep[$z=0.55$;][]{Chang2019A&A}  has the highest energy photon $\sim148$ GeV and lies below the horizon. Additional, Fermi-LAT, with its limited sensitivity, detected $>$1 TeV photons from five known VHE sources (J0349.4$-$1159, J1104.4$+$3812, J1653.8$+$3945, J2000.0$+$6508, J2158.8$-$3013). In Figure~\ref{fig:energy_redhsift_tauEBL}, a few sources appear to be located with EBL optical depths exceeding $\tau_{EBL} >\,2$. These opacities are evaluated corresponding to the highest energy photon detected from the direction of source.  These sources, though few, may provide valuable constraints on the $\gamma$-ray opacity of the Universe, and we briefly comment on them below. For comparison, the 1CGH work reports an EBL optical depth computed at the mean energy of the four highest-energy photons ($\langle \tau_{EBL} \rangle$; \citealt{Arsioli2025MNRAS}). The corresponding $\langle \tau_{EBL} \rangle$ values for the sources discussed here are also listed for reference.

{\it J0035.9$+$5950 (1ES 0033$+$595)}:
This object is a known VHE emitter with a synchrotron peak frequency of $\log \nu^{\rm peak}_{\rm syn}=18.2$ (in Hz). Two spectroscopic redshift measurements are reported: $z=0.467$ \citep{Paiano2017ApJ} and $z=0.086$ \citep{Michael2022ApJS}. In our work, we adopted $z=0.467$ since this redshift measurement was based on a higher signal-to-noise ratio optical spectrum \citep{Paiano2017ApJ}. Fermi-LAT registered 29 photons above 100 GeV, with the highest-energy photon of 655.3 GeV corresponding to an EBL optical depth of $\tau_{\rm EBL} = 4.352$. A $\langle \tau_{EBL} \rangle$ = 2.18  was reported in 1CGH catalog \citep{Arsioli2025MNRAS}.

{\it J0349.4$-$1159 (1ES 0347$-$121)}:
This extreme blazar is a known VHE emitter at $z=0.19$ \citep[][]{Arsioli2025MNRAS} and $\log \nu^{\rm peak}_{\rm syn}=18$ (in Hz). The energy of the highest energy photon detected by the Fermi$-$LAT is 1.25 TeV, corresponding to an EBL optical depth of $\tau_{\rm EBL} = 2.24$ ($\langle \tau_{EBL} \rangle=1.00$ reported in \citet{Arsioli2025MNRAS}).

{\it J0507.9$+$6737 (1ES 0502$+$675)}:
This object is also a known VHE emitting BL Lac object at the redshift of $z=0.34$ \citep[][]{2013ApJ...764..135S} and $\log \nu^{\rm peak}_{\rm syn}=17.9$ (in Hz). The energy of the highest-energy photon is 609.5 GeV, corresponding to an EBL optical depth $\tau_{\rm EBL} = 2.75$ ($\langle \tau_{EBL} \rangle = 2.28 $ reported by \citet{Arsioli2025MNRAS}). Our analysis yielded a test statistic of 433, with a total of 37 photons detected above 100 GeV.

{\it J2055.4$-$0020 (1RXS J205528.2$-$002123)}:
Identified as a new VHE emitting blazar for the first time in this work, this object is located at $z=0.44$ \citep[][]{2013ApJ...764..135S}. The source has $\log \nu^{\rm peak}_{\rm syn}=18$ (in Hz) and three photons were detected above 100 GeV at $>$95\% association probability, with the highest energy photon of 357 GeV yielding an optical depth of 2.082 which is in contrast with $\langle \tau_{EBL} \rangle = 0.56$ reported in 1CGH catalog \citep{Arsioli2025MNRAS}.

\subsection{Comparing with other catalogs}
Recently, \citet{Neronov2025arXiv} reported a catalog of VHE emitting AGN candidates at high Galactic latitudes using a photon-counting clustering technique, identifying a total of 175 high-confidence candidates, which include 71 previously known sources, and 127 HSP blazars. In contrast, we employed the full likelihood optimization technique to determine the detection significance and spectral parameters and found 144 VHE emitting HSP blazars. Among these, 106 objects are common in both works, including 42 established IACT-detected VHE-emitting blazars. The remaining 21 objects reported as VHE emitters by \citet{Neronov2025arXiv} have $9<{\rm TS}<25$ as per Fermi-LAT data analysis. 
\\
Additionally, \citet{Arsioli2025MNRAS} performed a $\gamma$-ray search in the recently released 1CGH catalog, identifying a total of 954 HSP blazars detected above 10 GeV. A complementary cross-match between our VHE catalog and the 1CGH sample reveals that approximately 15\% of $\gamma$-ray–emitting HSPs above 10 GeV exhibit significant VHE emission.

\begin{figure}
    \centering
    \includegraphics[width=\linewidth]{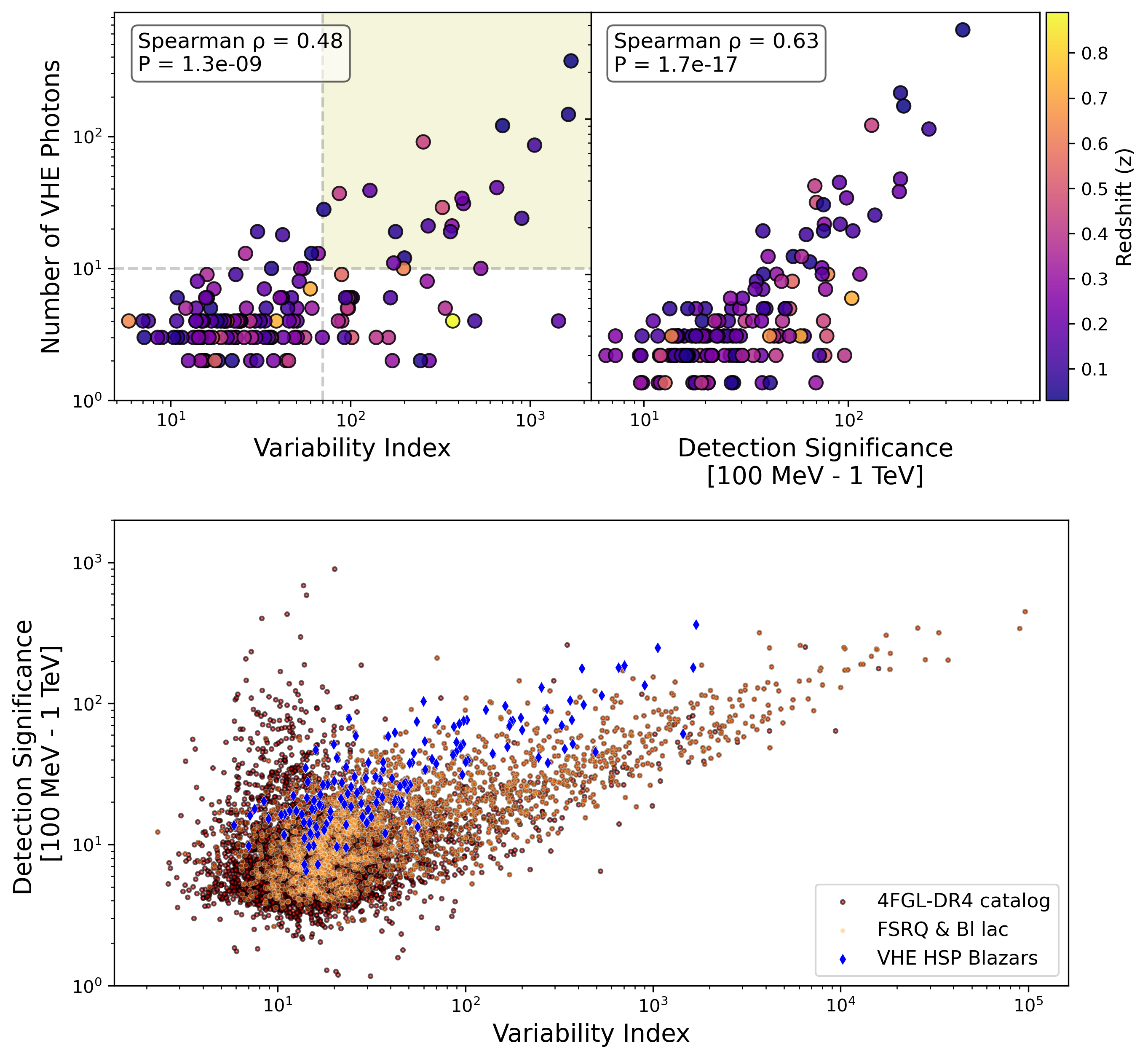}
    \caption{Top left: Dependence of number of VHE photons detected and the variability index. Top right: Dependence of number of VHE photons and detection significance. Bottom: Variability Index and its dependence on detection significance of the sources in 4FGL-DR4 catalog overlaid with FSRQ and Bl lac values and VHE sources identified in this work.}
    \label{fig:variability_plots_correlation}
\end{figure}

\subsection{Source Variability and VHE Time of arrival} 

\begin{figure*}
    \includegraphics[width=0.9\textwidth]{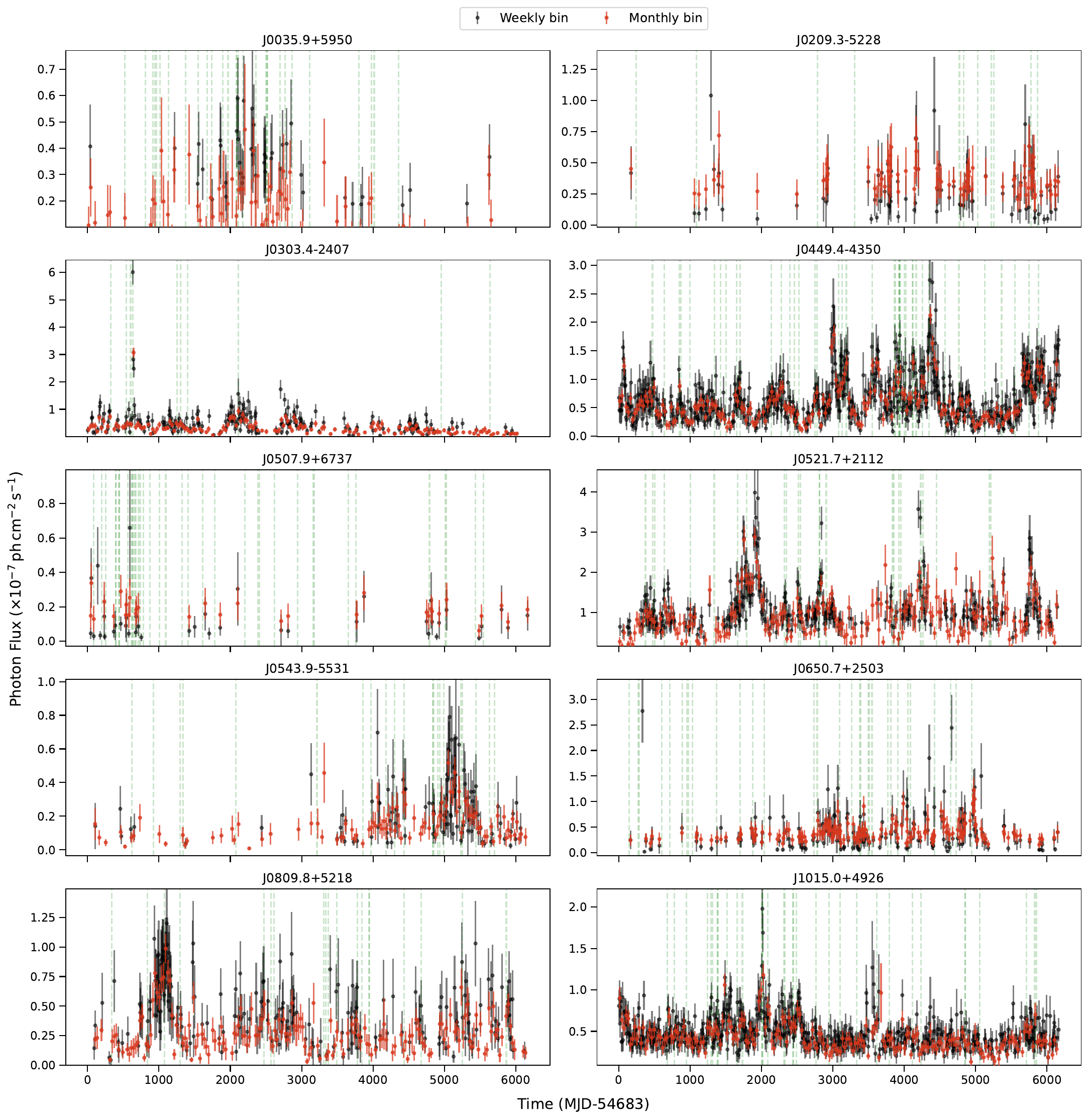}
    \caption{Weekly (black) and Monthly (red) binned Fermi-LAT lightcurve (100 MeV - 100 GeV) of sources highlighted in Top left panel of Figure \ref{fig:variability_plots_correlation} with a variability index $>$70 and atleast 10 VHE photon amongst the known and newly identified VHE sources presented in this work (Table \ref{tab:new_candidates_1} and \ref{tab:known_VHE_candidates}). Only significantly detected flux bins with TS$>$9 are plotted. The time of arrival of VHE photons for each source is highlighted with vertical green dashed lines.}
    \label{fig:All_Sources_Lightcurves_1}
\end{figure*}

\begin{figure*}
    \figurenum{5}
    \vbox{
    \includegraphics[width=0.9\textwidth]{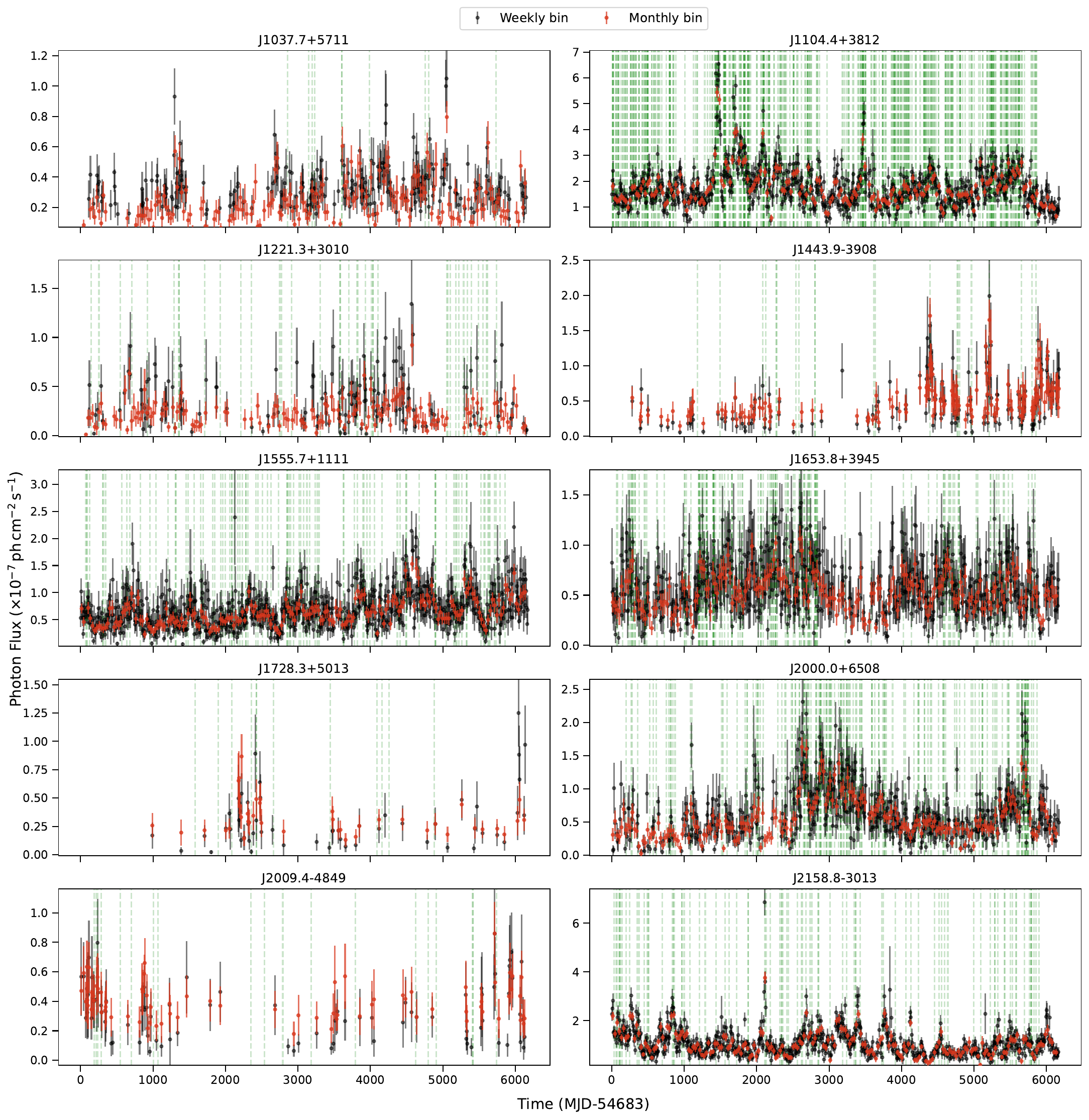}
    \includegraphics[width=0.45\textwidth]{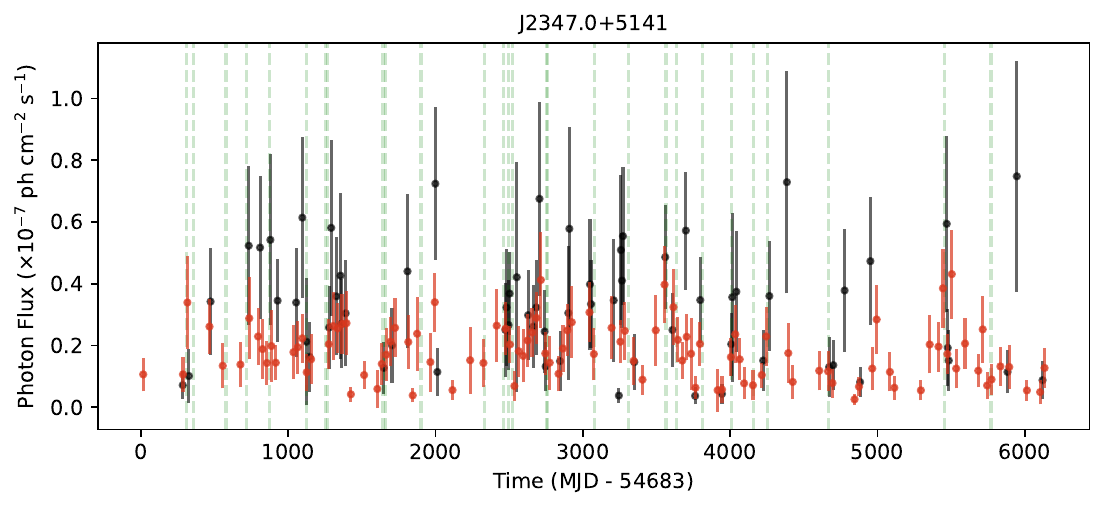}
        }
    \caption{Continued.}
\end{figure*}

Variability provides a crucial diagnostic of the emission processes in blazars, particularly in the VHE domain where photons are expected to originate from compact regions during episodes of enhanced particle acceleration \citep[e.g.,][]{2007ApJ...664L..71A}. Since the EGRET era, spectral hardening during flares has highlighted the enhanced detectability of VHE photons in such states. However, the observational strategies of current ground$-$based VHE facilities - typically triggered by flares - leave it uncertain whether VHE emission is confined to active states or also present at lower, quiescent flux levels. The latter possibility was proposed by \citet{Stecker_salmon1996ApJ} to account for the diffuse extragalactic $\gamma-$ray background (\citealt{Fichtel1996A&AS}; \citealt{Sreekumar1998}), and has been supported by dedicated observations such as the extended low-state campaign of PKS 2155$–$304 from 2005$–$2007 \citep{HESS2010A&A}. Yet, such detections remain limited by sensitivity and the long exposures required.

With Fermi-LAT, it is possible to probe the VHE domain continuously, independent of flaring triggers. In our analysis, we find that the number of detected VHE photons shows a positive correlation with both the source variability index and its overall detection significance, with a Spearman correlation coefficient of 0.48 (Fig. \ref{fig:variability_plots_correlation}). This suggests that sources exhibiting stronger variability tend to yield more VHE photons, consistent with the expectation that highly variable systems undergo more frequent episodes of efficient particle acceleration. At the same time, the variability index and detection significance are themselves tightly correlated, reflecting the fact that brighter sources are more likely to exhibit measurable variability. Disentangling the roles of intrinsic variability and observational bias is therefore critical. We also note that an apparent upper envelope exists in the number of VHE photons at a given variability index. 

To investigate whether VHE photon arrival is preferentially associated with active source states, we examined the weekly and monthly binned light curves of sources with the 4FGL-DR4 variability index $>70$ and at least 10 detected VHE photons (Fig. \ref{fig:All_Sources_Lightcurves_1}). The lightcurves are used from Fermi-LAT lightcurve repository \footnote{\url{https://fermi.gsfc.nasa.gov/ssc/data/access/lat/LightCurveRepository/}} and are in energy range from 100 MeV$-$100 GeV. In these plots, the arrival times of individual VHE photons are marked as vertical green dashed lines.

We find that while several VHE photons coincide with pronounced high-energy flaring episodes, e.g., in 4FGL~J1104.4+3812 or Mrk~421, a considerable fraction are instead detected during epochs consistent with the blazar being in a low activity. This indicates that VHE emission is not exclusively tied to strong flares, but can also persist at lower flux states. The presence of such quiescent VHE photons may reflect a steady acceleration component in the jet consistent with earlier claims of baseline VHE activity in blazars \citep{HESS2010A&A,Dmytriiev2021MNRAS}. Conversely, the clustering of photon arrivals during flares reinforces the link between efficient acceleration of energetic particles and enhanced VHE production.

These results suggest that both transient flares and underlying quiescent processes contribute to the observed VHE emission, highlighting the need for continuous monitoring to disentangle their relative roles.

\subsection{Prospects for followup with Cherenkov Telescopes}
\begin{figure}
\centering
\includegraphics[scale=0.4]{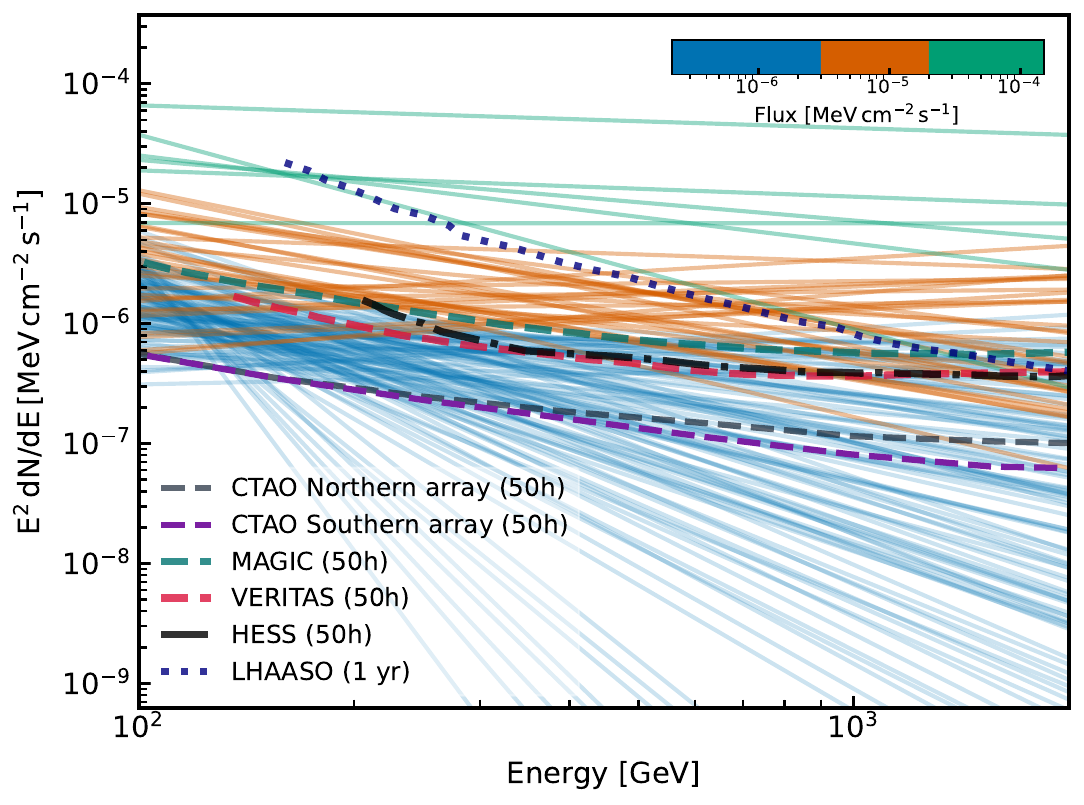}
\caption{Very-high-energy $\gamma$-ray spectra of significant sources identified in this work, color-coded by flux levels as defined in the inset. Overlaid are the differential sensitivity curves of CTAO, MAGIC, VERITAS, H.E.S.S. for an observation time of 50 hr and for LHAASO for 1 year.}
\label{fig:VHE_telescope_sensitivity}
\end{figure}

The newly identified VHE emitting blazars span a flux range covering approximately two orders of magnitude in the 100 GeV$–$2 TeV energy band. To characterize their distribution, we divided the total flux range into three bins:
\begin{enumerate}
\item \textbf{Low flux:} 108 sources with fluxes $<3 \times 10^{-6}\,\mathrm{MeV\,cm^{-2}\,s^{-1}}$,
\item \textbf{Intermediate flux:} 30 sources with fluxes in the range $(3$–$20) \times 10^{-6}\,\mathrm{MeV\,cm^{-2}\,s^{-1}}$,
\item \textbf{High flux:} 6 sources with fluxes in the range $(2$–$15) \times 10^{-5}\,\mathrm{MeV\,cm^{-2}\,s^{-1}}$.
\end{enumerate}

In Figure~\ref{fig:VHE_telescope_sensitivity}, we show the 0.1$-$2 TeV power law spectrum of all sources and overplot the sensitivity limits of several Cherenkov telescopes. As can be seen, some blazars present in the intermediate- and high-flux categories should be easily detectable with currently operating Cherenkov facilities - MAGIC, VERITAS, H.E.S.S and LHAASO. Furthermore,  a major fraction of all sources should be detectable with the improved sensitivity of the upcoming CTAO.

\section{Summary}    \label{sec:summary}
In this work, we analyzed the 0.1$-$2 TeV Fermi-LAT data covering $\sim$16 years of the satellite operation, to identify VHE emitting HSP blazars. We summarize our findings below:

\begin{itemize}
    \item We identified 92 new VHE-emitting HSP blazars for the first time, detected at a high confidence level (TS $\geq$ 25, equivalent to $\geq$ 5$\sigma$). The study also confirmed VHE emission from 52 previously known VHE blazars listed in the TeVCat catalog. These sources constitute excellent targets for follow-up observations with currently operating Cherenkov facilities - MAGIC, VERITAS and H.E.S.S., and LHAASO - as well as the forthcoming CTAO. Furthermore, an additional 116 sources were detected at moderate significance (12 $\leq$ TS $<$ 25) and can be regarded as promising VHE candidates for future observations.

    \item A comparison with non-VHE emitting HSP blazars revealed that the identified VHE sources are generally located at lower redshifts (mean $z = 0.2 \pm 0.1$), and exhibit brighter synchrotron emission ($\log F^{\rm peak}_{\rm syn} = -11.22 \pm 0.45$, in erg cm$^{-2}$ s$^{-1}$). These characteristics suggest a potential Malmquist bias, where brighter and closer objects are more easily detected.

    \item VHE blazars stand out from the general $\gamma$-ray population by their hard spectra and moderate brightness. Interestingly, VHE blazars tend to exhibit relatively low-to-moderate variability indices for their flux level, consistent with expectations from spectral considerations: a hard LAT spectrum ($\Gamma<2$) implies an inverse Compton peak at hundreds of GeV, where the LAT band is dominated by lower-energy electrons with longer cooling times, naturally leading to reduced variability compared to the TeV regime.\\

    \item We found that the number of detected VHE photons correlates positively with both the variability index and detection significance, consistent with more variable (and brighter) systems undergoing frequent episodes of efficient particle acceleration. However, this correlation is partly influenced by observational bias, as brighter sources are easier to detect and characterize. By examining the arrival times of VHE photons against source light curves of strongly variable sources, the study shows that while many VHE photons arrive during major flares, a significant fraction are also detected during quiescent states. This dual behavior indicates that VHE emission is not confined to flaring episodes, but can also arise from a steady, baseline acceleration component in jets. \\

\end{itemize}

\section{Software and third party data repository citations} \label{sec:cite}

\facilities{$Fermi$-LAT}

\software{numpy \citep{Taylor2005_TOPCAT},  
          matplotlib \citep{Hunter:2007}, 
          TopCAT\citep{harris2020array}, Fermipy\citep{Wood:2017TJ}
          }




\appendix
In Table~\ref{tab:ts_betwen_12_and_25}, we provide the spectral parameters of HSP blazars detected with TS between 12 and 25. The remaining HSP blazars with TS$<12$ and with at least one VHE photon associated with more than 95\% probability of association with the sources , are tabulated in Table~\ref{tab:ts_less_than_9}. Table \ref{tab:unassociated_sources} summarizes the list of unassociated Fermi $\gamma$-ray sources that exhibit atleast one VHE photons. Table \ref{tab:ultraClean_class_ts_lessthan25} lists detection statistics of sources with different event class cuts -- SOURCE, CLEAN and ULTRACLEAN.

\begin{deluxetable*}{lccccccc}
\tabletypesize{\footnotesize}
\tablecaption{Sources with TS between 12 and 25 \label{tab:ts_betwen_12_and_25}}
\tablewidth{0pt}
\tablehead{
\colhead{Source Name} & 
\colhead{$\rm TS_{\rm obs}$} & 
\colhead{F$_{\rm obs}$} & 
\colhead{$\Gamma_{\rm obs}$} &
\colhead{index\_flag} &
\colhead{No.} & 
\colhead{E$_{\rm max}$} & 
\colhead{$\rm Time_{\rm arr}$} \\
\colhead{[1]} & 
\colhead{[2]} & 
\colhead{[3]} & 
\colhead{[4]} & 
\colhead{[5]} & 
\colhead{[6]} & 
\colhead{[7]} & 
\colhead{[8]} 
}
\startdata
J0013.9$-$1854 & 16 & $0.71 \pm 0.68$ & $2.66 \pm 1.37$ & \quad 0  & 2 & 314.6 & 58670.1 \\
J0039.1$-$2219 & 21 & $1.49 \pm 0.96$ & $2.75 \pm 0.95$ & \quad 0  & 2 & 309.1 & 57207.6 \\
J0040.3$+$4050 & 18 & $1.05 \pm 0.95$ & $2.42 \pm 1.07$ & \quad 0  & 2 & 159.1 & 60008.0 \\
J0045.3$+$2128 & 17 & $1.10 \pm 1.07$ & $2.18 \pm 1.03$ & \quad 0  & 2 & 246.8 & 54998.7 \\
J0047.9$+$3947 & 13 & $0.81 \pm 0.70$ & $2.52 \pm 1.14$ & \quad 0  & 2 & 292.2 & 58842.6 \\
\enddata
\tablecomments{The column descriptions are identical to those in Table~\ref{tab:new_candidates_1}. The index\_flag in column [5] is set to 1 for sources with $\gamma_{\mathrm{obs}} = 10.0$ in column [4], and 0 otherwise. The complete version of this table is available on ZENODO (doi: \href{https://doi.org/10.5281/zenodo.17593405}{10.5281/zenodo.17593405}).}
\end{deluxetable*}

\begin{table}
\caption{The remaining HSP blazars with TS less than $12$ and at least one VHE photon detected with more than 95\% probability of association. \label{tab:ts_less_than_9}}
\begin{tabular}{lccc}
\hline
Source Name & No. & E$_{\rm max}$ & Time$_{\rm arr}$ \\
$[1]$ & $[2]$ & $[3]$ & $[4]$ \\
\hline
J0018.4+2946  & 1 & 127.3 & 56743.5 \\
J0021.9-5140  & 1 & 194.8 & 60513.5 \\
J0033.3-2040  & 1 & 219.7 & 57500.7 \\
J0035.2+1514  & 1 & 264.0 & 57134.3 \\
J0056.3-0935  & 2 & 310.4 & 57177.7 \\
\hline
\end{tabular}
\tablecomments{The column information are same as in Table~\ref{tab:new_candidates_1}. The complete table is available at ZENODO (doi: \href{https://doi.org/10.5281/zenodo.17593405}{10.5281/zenodo.17593405}).}
\end{table}

\begin{deluxetable}{lcccc}
\tablecaption{Unassociated Fermi sources from with at least one VHE photon was detected. \label{tab:unassociated_sources}}
\tablehead{
\colhead{Source Name} &
\colhead{No.} & 
\colhead{E$_{\rm max}$} & 
\colhead{$\rm Time_{\rm arr}$} \\
\colhead{[1]} & 
\colhead{[2]} & 
\colhead{[3]} & 
\colhead{[4]}  
}
\startdata
J0057.9$+$6326   & 1 & 249.4 & $54922.0$ \\
J0357.7$-$6808   & 1 & 110.6 & $54687.0$ \\
J0438.0$-$7329   & 1 & 155.1 & $58435.8$ \\
J1146.0$-$0638   & 1 & 113.4 & $56170.9$ \\
J1452.0$-$4148   & 1 & 346.3 & $57104.2$ \\
\enddata
\tablecomments{The column descriptions are identical to those in Table~\ref{tab:new_candidates_1}. The complete version of this table is available on ZENODO (doi: \href{https://doi.org/10.5281/zenodo.17593405}{10.5281/zenodo.17593405}).}
\end{deluxetable}

\begin{deluxetable*}{lcccccccc}
\tablewidth{0pt}
\tablecaption{Sources in Table \ref{tab:new_candidates_1} 
 and \ref{tab:known_VHE_candidates} with TS $<$ 25 with CLEAN and ULTRACLEAN class event cuts \label{tab:ultraClean_class_ts_lessthan25}. \label{tab:Ultraclean_class}}
\tablehead{
\colhead{Source name} &
\colhead{$\rm{TS_{SOURCE}}$} & 
\colhead{$\rm{TS_{evclass}}$} &
\colhead{$\rm F_{\rm obs,SOURCE}$} & 
\colhead{$\rm F_{\rm obs,evclass}$} & 
\colhead{$\rm No._{\mathrm{SOURCE}}$} & 
\colhead{$\rm No._{\mathrm{evclass}}$} \\ 
\colhead{[1]} & 
\colhead{[2]} & 
\colhead{[3]} & 
\colhead{[4]} & 
\colhead{[5]} & 
\colhead{[6]} & 
\colhead{[7]}
}
\startdata
\hline
\multicolumn{7}{|c|}{\textbf{Sources with TS$<$ 25 in both CLEAN and ULTRACLEAN class event cuts.}}\\
\multicolumn{7}{|c|}{\textbf{Values listed here are from evclass$=$CLEAN class}}\\
\hline
J0516.4$+$7350$^*$ & 25 & 8  & $0.92 \pm 0.68$ & $0.47 \pm 0.62$ & 3 & 1 \\
J0934.5$-$1720$^*$ & 25 & 23 & $0.79 \pm 0.51$ & $0.77 \pm 0.53$ & 3 & 3 \\
J1130.5$-$7801$^*$ & 38 & 16 & $0.85 \pm 0.44$ & $0.61 \pm 0.43$ & 4 & 2 \\
J1813.5$+$3144$^*$ & 26 & 13 & $0.97 \pm 0.59$ & $0.96 \pm 1.02$ & 4 & 2 \\
J1954.9$-$5640$^*$ & 26 & 22 & $1.02 \pm 0.70$ & $1.10 \pm 0.79$ & 2 & 2 \\
J2026.1$+$7645$^*$ & 26 & 17 & $0.82 \pm 0.58$ & $0.53 \pm 0.46$ & 3 & 2 \\
J2001.2$+$4353$^*$ & 38 & 13 & $0.88 \pm 0.45$ & $0.66 \pm 0.40$ & 5 & 5 \\
\hline
\multicolumn{7}{|c|}{\textbf{Sources with TS$<$ 25 in evclass $=$ CLEAN class event cuts only}}\\
\hline
J0014.7$+$5801 & 27 & 19 & $0.77 \pm 0.50$ & $0.69 \pm 0.54$ & 3 & 2 \\
J0316.2$+$0905 & 29 & 24 & $2.00 \pm 1.42$ & $0.98 \pm 0.69$ & 4 & 3 \\
J0353.0$-$6831 & 35 & 21 & $1.55 \pm 0.86$ & $1.14 \pm 0.81$ & 3 & 3 \\
J1944.0$+$2117 & 51 & 23 & $5.42 \pm 2.59$ & $3.00 \pm 1.76$ & 6 & 5 \\
\hline
\multicolumn{7}{|c|}{\textbf{Sources with TS$<$ 25 in evclass $=$ ULTRACLEAN class event cuts only}}\\
\hline
J0014.7$+$5801 & 27 & 19 & $0.77 \pm 0.50$ & $0.79 \pm 0.64$ & 3 & 2 \\
J0015.6$+$5551 & 39 & 15 & $1.57 \pm 0.83$ & $0.84 \pm 0.66$ & 4 & 2 \\
J0349.4$-$1159 & 43 & 20 & $3.29 \pm 2.09$ & $3.59 \pm 3.63$ & 5 & 2 \\
J0350.0$+$0640 & 31 & 23 & $0.49 \pm 0.37$ & $0.25 \pm 0.18$ & 3 & 2 \\
J0353.0$-$6831 & 35 & 22 & $1.55 \pm 0.86$ & $1.34 \pm 1.00$ & 3 & 3 \\
J0744.1$+$7434 & 26 & 10 & $0.68 \pm 0.41$ & $0.46 \pm 0.39$ & 3 & 1 \\
J0747.5$-$4927 & 25 & 21 & $2.15 \pm 1.53$ & $2.90 \pm 2.49$ & 2 & 1 \\
J1353.6$-$6640 & 51 & 21 & $1.35 \pm 0.63$ & $0.80 \pm 0.53$ & 5 & 3 \\
J1506.6$+$0813 & 27 & 22 & $0.45 \pm 0.33$ & $0.41 \pm 0.34$ & 3 & 2 \\
J1744.0$+$1935 & 29 & 18 & $0.95 \pm 0.65$ & $0.62 \pm 0.59$ & 3 & 2 \\
J1911.4$-$1908 & 25 & 23 & $0.96 \pm 0.57$ & $1.03 \pm 0.77$ & 2 & 2 \\
J2343.6$+$3438 & 35 & 23 & $1.24 \pm 0.67$ & $0.86 \pm 0.66$ & 3 & 2 \\
\enddata
\tablecomments{The column information are as follows: [1]: 4FGL source name. Sources with $^*$ did not cross CLEAN and ULTRACLEAN class cut both; [2]: TS with SOURCE class cut [3]: TS with CLEAN or ULTRACLEAN class cut as specified, [4] and [5]: energy flux (in 10$^{-6}$ MeV cm$^{-2}$ s$^{-1}$) in the energy range of 0.1$-$2 TeV with SOURCE and evclass events cut (CLEAN or ULTRACLEAN) as specified; [6] and [7] number of $>$100 GeV photons detected from the source having $\geq$95\% association probability with SOURCE and evclass events  cut (CLEAN or ULTRACLEAN) as specified.}
\end{deluxetable*}

\bibliography{main}{}
\bibliographystyle{aasjournal}

\end{document}